\documentclass{article}
\usepackage{spconf,amsmath,epsfig,graphicx,color}
\graphicspath{{Eps/}}

\input{alphabet.tex}
\def\bdoc{\begin{document}}             \def\edoc{\end{document}}
\def\babs{\begin{abstract}}             \def\eabs{\end{abstract}}
\def\bcc{\begin{center}}                \def\ecc{\end{center}}
\def\ben{\begin{enumerate}}             \def\een{\end{enumerate}}
\def\bit{\begin{itemize}}               \def\eit{\end{itemize}}
\def\bpic{\begin{picture}}              \def\epic{\end{picture}}
\def\beq{\begin{equation}}              \def\eeq{\end{equation}}
\def\barr{\begin{array}}                \def\earr{\end{array}}
\def\btab{\begin{tabular}}              \def\etab{\end{tabular}}
\def\btabu{\begin{tabular}}             \def\etabu{\end{tabular}}
\def\bcc{\begin{center}}                \def\ecc{\end{center}}
\def\beqn{\begin{eqnarray}}             \def\eeqn{\end{eqnarray}}
\def\beqnx{\begin{eqnarray*}}           \def\eeqnx{\end{eqnarray*}}
\def\bfig{\begin{figure}}               \def\efig{\end{figure}}
\def\bver{\begin{verb}}						 \def\ever{\end{verb}}

\def\d#1{\,\mbox{d} #1}
\def\disp#1{\displaystyle{#1}}

\def\iii{\int_{-\infty}^{+\infty}}
\def\izi{\int_{0}^{\infty}}
\def\izpi{\int_{0}^{\pi}}
\def\izdpi{\int_{0}^{2\pi}}
\def\intd{\int\kern-.8em\int}
\def\intt{\int\kern-.8em\int\kern-.8em\int}
\def\intg{\kern-1.1em\int}

\def\xh{\widehat{x}}
\def\thetah{\widehat{\theta}}
\def\betah{\widehat{\beta}}

\def\xbh{\widehat{\xb}}
\def\thetabh{\widehat{\thetab}}
\def\betabh{\widehat{\betab}}

\def\xbhk{\widehat{\xb}^{k}}
\def\thetahk{\widehat{\theta}^{k}}
\def\betahk{\widehat{\beta}^{k}}

\def\xbhkp{\widehat{\xb}^{k+1}}
\def\thetahkp{\widehat{\theta}^{k+1}}
\def\betahkp{\widehat{\beta}^{k+1}}

\def\thetabhk{\widehat{\thetab}^{k}}
\def\betabhk{\widehat{\betab}^{k}}

\def\thetabhkp{\widehat{\thetab}^{k+1}}
\def\betabhkp{\widehat{\betab}^{k+1}}

\def\thetamin{\theta_{\mbox{\tiny min}}}
\def\thetamax{\theta_{\mbox{\tiny max}}}
\def\betamin{\beta_{\mbox{\tiny min}}}
\def\betamax{\beta_{\mbox{\tiny max}}}

\def\ra{\rightarrow}
\def\la{\leftarrow}
\def\da{\downarrow}
\def\ua{\uparrow}

\def\Ra{\Rightarrow}
\def\La{\Leftarrow}
\def\Da{\Downarrow}
\def\Ua{\Uparrow}

\def\lra{\longrightarrow}
\def\lla{\longleftarrow}
\def\Lra{\Longrightarrow}
\def\Lla{\longleftarrow}

\def\lrarr{\leftrightarrow}
\def\Lrarr{\Leftrightarrow}
\def\udarr{\updownarrow}
\def\Uparr{\Updownarrow}

\def\expf#1{\exp\left[ {#1} \right]}

\def\argmax#1#2{\arg\max_{#1}\left\{ #2 \right\}}
\def\argmin#1#2{\arg\min_{#1}\left\{ #2 \right\}}

\def\Prob#1{\mbox{Pr}\left\{#1\right\}}
\def\var#1{\mbox{Var}\left\{#1\right\}}
\def\cov#1{\mbox{Cov}\left\{#1\right\}}
\def\corr#1{\mbox{Corr}\left\{#1\right\}}
\def\trace#1{\mbox{Tr}\left\{#1\right\}}
\def\rang#1{\mbox{rang}\left\{#1\right\}}
\def\det#1{\mbox{d\'et}\left\{#1\right\}}

\def\gbu{\underline{\gb}}
\def\fbu{\underline{\fb}}
\def\zbu{\underline{\zb}}
\def\epsilonbu{\underline{\epsilonb}}
\def\thetabu{\underline{\thetab}}

\def\sigmae{{\sigma_{\epsilon}}}
\def\Sigmae{{\Sigmab_{\epsilonb}}}
\def\Sigmaf{{\Sigmab_{\fb}}}
\def\Sigmag{{\Sigmab_{\gb}}}

\def\fbuh{\widehat{\fbu}}
\def\zbuh{\widehat{\fbu}}
\def\thetabuh{\widehat{\thetabu}}

\def\fbh{\widehat{\fb}}
\def\Pbh{\widehat{\Pb}}
\def\Sigmabh{\widehat{\Sigmab}}
\def\Abh{\widehat{\Ab}}
\def\Bbh{\widehat{\Bb}}
\def\thetabh{\widehat{\thetab}}
\def\Rbe{\Rb_{\epsilon}}
\def\Rbeh{\widehat{\Rbe}}

\def\Rjk{{R_j}_k}
\def\fbik{{\fb_i}_k}
\def\fbjk{{\fb_j}_k}

\def\mik{{m_i}_k}
\def\sigmaik{{\sigma_i^2}_k}
\def\Sigmaik{{\Sigmab_i}_k}
\def\alphaik{{\alpha_i}_k}
\def\betaik{{\beta_i}_k}

\def\muik{{\mu_i}_k}
\def\vik{{v_i}_k}

\def\mjk{{m_j}_k}
\def\sigmajk{{\sigma_j^2}_k}
\def\Sigmajk{{\Sigmab_j}_k}
\def\alphajk{{\alpha_j}_k}
\def\betajk{{\beta_j}_k}

\def\mujk{{\mu_j}_k}
\def\vjk{{v_j}_k}
\def\njk{{n_j}_k}

\def\alphae{\alpha_{\epsilon}}
\def\betae{\beta_{\epsilon}}
\def\alphaez{\alpha_{\epsilon_0}}
\def\betaez{\beta_{\epsilon_0}}
\def\alphaei{\alpha_{\epsilon_i}}
\def\betaei{\beta_{\epsilon_i}}
\def\alphaeiz{\alpha_{\epsilon_{i_0}}}
\def\betaeiz{\beta_{\epsilon_{i-0}}}

\def\mbz{{{\mb}_z}}
\def\Sigmabz{{{\Sigmab}_z}}
\def\diag#1{\mbox{diag}\left[#1\right]}

\def\sigmah{\widehat{\sigma}}
\def\fh{\widehat{f}}
\def\sigmaz{\sigma_z}

\def\blue#1{{\color{blue}#1}}
\def\red#1{{\color{red}#1}}
\def\green#1{{\color{green}#1}}

\def\sumi{\sum_{i=1}^{M} }
\def\sumj{\sum_{j=1}^{N} }
\def\lambdah{\wh{\lambda}}
\def\lambdabh{\wh{\lambdab}}

\def\det#1{\hbox{det}\left(#1\right)}
\def\defined{\,\shortstack{$\triangle$\\ =}\,}

\def\zmat{\left[\matrix{0}\right]}
\def\det#1{\hbox{det}(#1)}
\def\th{\widehat{t}}
\def\xh{\widehat{x}}
\def\lambdah{\widehat{\lambda}}
\def\muh{\widehat{\mu}}
\def\sigmah{\widehat{\sigma}}
\def\rhoh{\widehat{\rho}}
\def\betah{\widehat{\beta}}
\def\alphah{\widehat{\alpha}}
\def\tbh{\widehat{\tb}}
\def\mubh{\widehat{\mub}}
\def\sigmabh{\widehat{\sigmab}}
\def\rhobh{\widehat{\rhob}}
\def\oneb{\mbox{\bf 1}}

\def\bm#1{\mbox{\boldmath $#1$}}
\def\zerob{{\bf 0}}
\def\oneb{{\bf 1}}
\def\intg{\int\kern-1.1em\int}
\def\expf#1{\exp\left[ {#1} \right]}

\def\gbu{\underline{\gb}}
\def\fbu{\underline{\fb}}
\def\zbu{\underline{\zb}}
\def\epsilonbu{\underline{\epsilonb}}
\def\uthetab{\underline{\thetab}}

\def\sigmae{{\sigma_{\epsilon}}}
\def\Sigmae{{\Sigma_{\epsilonb}}}
\def\Sigmabe{{\Sigmab_{\epsilonb}}}
\def\Sigmaf{{\Sigmab_{\fb}}}
\def\Sigmag{{\Sigmab_{\gb}}}

\def\fbuh{\widehat{\fbu}}
\def\zbuh{\widehat{\fbu}}
\def\uthetabh{\widehat{\uthetab}}

\def\fbh{\widehat{\fb}}
\def\Sigmabh{\widehat{\Sigmab}}
\def\Abh{\widehat{\Ab}}
\def\thetabh{\widehat{\thetab}}

\def\Rjk{{R_j}_k}
\def\fbik{{\fb_i}_k}
\def\fbjk{{\fb_j}_k}

\def\mik{{m_i}_k}
\def\sigmaik{{\sigma_i^2}_k}
\def\Sigmaik{{\Sigmab_i}_k}
\def\alphaik{{\alpha_i}_k}
\def\betaik{{\beta_i}_k}

\def\muik{{\mu_i}_k}
\def\vik{{v_i}_k}

\def\mjk{{m_j}_k}
\def\sigmajk{{\sigma_j^2}_k}
\def\Sigmajk{{\Sigmab_j}_k}
\def\alphajk{{\alpha_j}_k}
\def\betajk{{\beta_j}_k}

\def\mujk{{\mu_j}_k}
\def\vjk{{v_j}_k}
\def\njk{{n_j}_k}

\def\mbz{{{\mb}_z}}
\def\Sigmabz{{{\Sigmab}_z}}

\def\vect{\mbox{vect}}
\def\lra{\longrightarrow}

\def\gir{g_i(\rb)}
\def\fjr{f_j(\rb)}
\def\zjr{z_j(\rb)}
\def\epsilonir{\epsilon_i(\rb)}
\def\gbr{\gb(\rb)}
\def\fbr{\fb(\rb)}
\def\zbr{\zb(\rb)}
\def\epsilonbr{\epsilonb(\rb)}
\def\fbhr{\fbh(\rb)}
\def\Sigmabhr{\Sigmabh(\rb)}

\def\mbzr{\mb_{z(\rb)}}
\def\Sigmabzr{\Sigmab_{z(\rb)}}
\def\Sigmagbzr{{\Sigmab_g}_{z(\rb)}}

\def\mjzjr#1{{m_{#1}}_{z_{#1}(\rb)}}
\def\sigmajzjr#1{{\sigma_{#1}}_{z_{#1}(\rb)}}

\def\Beta{B}
\def\Betab{\bm{\Beta}}
\def\Betabe{\bm{\Beta}_{\epsilonb}}

\def\fh{\widehat{f}}
\def\sigmah{\widehat{\sigma}}
\def\sigmaei{\sigma_{\epsilon_i}^2}

\def\thetah{\widehat{\theta}}
\def\sigmah{\widehat{\sigma}}

\def\Ftxm{\widetilde{F}_{X}(x_-)}
\def\Ftxp{\widetilde{F}_{X}(x_+)}

\def\NU{{\cal V}}

\def\tFx{\widetilde{F}_{X}}

\def\Fx{F_{X}(x)}
\def\fx{f_{X}(x)}

\def\Fxv{F_{X}(x)}
\def\fxv{f_{X}(x)}

\def\Fxi{{F}_{X_i}(x_i)}
\def\fxi{{f}_{X_i}(x_i)}

\def\Gnu{G_{\NU}(\nu)}
\def\gnu{g_{\NU}(\nu)}

\def\Fnu{F_{\NU}(\nu)}
\def\fnu{f_{\NU}(\nu)}

\def\Finnu#1{F^{-1}_{\NU}(#1)}

\def\Gnua#1{G_{\NU}(#1)}
\def\Gnub#1{G_{\NU}^{-1}(#1)}
\def\Gnuh{G_{\NU}^{-1}(1/2)}

\def\Ftx{\widetilde{F}_{X}(x)}
\def\ftx{\widetilde{f}_{X}(x)}

\def\Ftxi{\widetilde{F}_{X_i}(x_i)}
\def\ftxi{\widetilde{f}_{X_i}(x_i)}

\def\Ftxv{\widetilde{F}_{X}(x)}
\def\FtX#1{\widetilde{F}_{X}(#1)}
\def\ftxv{\widetilde{f}_{X}(x)}

\def\fxnu{f_{X|\NU}(x|\nu)}
\def\fxNU{f_{X|\NU}(x|\NU)}
\def\FxNU{F_{X|\NU}(x|\NU)}

\def\fxNUtheta{f_{X|\NU,\theta}(x|\NU,\theta)}
\def\FxNUtheta{F_{X|\NU,\theta}(x|\NU,\theta)}

\def\Fxnu{F_{X|\NU}(x|\nu)}
\def\Fxnui#1{F_{X|\NU}(x|\nu_#1)}

\def\FxN#1{F_{X|\NU}(#1)}
\def\FxNU{F_{X|\NU}(x|\NU)}
\def\Fxnun#1{F_{X|\NU}(x|#1)}
\def\Fxinun#1{F_{X_i|\NU}(x_i|#1)}
\def\fxnu{f_{X|\NU}(x|\nu)}
\def\fxNU{f_{X|\NU}(x|\NU)}

\def\Fxvnu{F_{X|\NU}(x|\nu)}
\def\Fxvnun#1{F_{X|\NU}(x|#1)}
\def\FxvNU{F_{X|\NU}(x|\NU)}

\def\FXvnu{F_{X|\NU}}

\def\fxvnu{f_{X|\NU}(x|\nu)}
\def\fxvNU{f_{X|\NU}(x|\NU)}

\def\FXvNU#1{{F}_{X|\NU}(#1|\NU)}
\def\FXvn#1{{F}_{X|\NU}(x|#1)}

\def\Fxtheta{F_{X|\theta}(x|\theta)}
\def\fxtheta{f_{X|\theta}(x|\theta)}

\def\Fxvtheta{F_{X|\theta}(x|\theta)}
\def\fxvtheta{f_{X|\theta}(x|\theta)}

\def\Ftxtheta{\widetilde{F}_{X|\theta}(x|\theta)}
\def\ftxtheta{\widetilde{f}_{X|\theta}(x|\theta)}

\def\Ftxvtheta{\widetilde{F}_{X|\theta}(x|\theta)}
\def\ftxvtheta{\widetilde{f}_{X|\theta}(x|\theta)}

\def\Fxnutheta{F_{X|\NU,\theta}(x|\nu,\theta)}
\def\fxnutheta{f_{X|\NU,\theta}(x|\nu,\theta)}

\def\ftheta{f_{\Theta}(\theta)}
\def\fthetaxnuz{f_{\Theta|X,\nu_0}(\theta|x,\nu_0)}
\def\fthetax{f_{\Theta|X}(\theta|x)}

\def\Fxvnutheta{F_{X|\NU,\theta}(x|\nu,\theta)}
\def\FxvNUtheta{F_{X|\NU,\theta}(x|\NU,\theta)}
\def\fxvnutheta{f_{X|\NU,\theta}(x|\nu,\theta)}
\def\Ftxvnutheta{\widetilde{F}_{X|\theta}(x|\theta)}
\def\ftxvnutheta{\widetilde{f}_{X|\theta}(x|\theta)}

\def\FxvNUm{F_{X|\NU}(x_-|\NU)}
\def\FxvNUp{F_{X|\NU}(x_+|\NU)}

\def\Fxinu{F_{X_i|\NU}(x_i|\nu)}
\def\fxinu{f_{X_i|\NU}(x_i|\nu)}

\def\Ftxi{\widetilde{F}_{X_i}(x_i)}
\def\ftxi{\widetilde{f}_{X_i}(x_i)}

\def\fxitheta{f_{X_i|\theta}(x_i|\theta)}
\def\ftxitheta{\widetilde{f}_{X_i|\theta}(x_i|\theta)}

\def\thetah{\widehat{\theta}}
\def\thetat{\widetilde{\theta}}

\def\lnuxv{l_{\nu|X}(\nu|x)}

\def\Lnuxva#1{L_{\nu|X}(#1 | x)}
\def\lnuxva#1{l_{\nu|X}(#1 | x)}

\def\Prob#1{\mbox{P}\left(#1\right)}

\def\lthetaxvnuz{l(\theta)}

\def\lthetaxvnuz{l_{\theta|x,\nu_0}(\theta | x,\nu_0)}
\def\fxnuztheta{f_{X|\nu_0,\theta}(x | \nu_0,\theta)}
\def\fxinuztheta{f_{X|\nu_0,\theta}(x_i | \nu_0,\theta)}
\def\fxvtheta{f_{X|\theta}(x | \theta)}
\def\lthetaxv{l_{\theta|X}(\theta | x)}
\def\ltthetaxv{\widetilde{l}_{\theta|X}(\theta | x)}

\def\ltheta{l(\theta )}
\def\lttheta{\widetilde{l}(\theta)}
\def\Lnu{L(\nu )}
\def\Lnui#1{L(\nu_#1)}
\def\Linu{L_i(\nu )}
\def\Linu{L_i(\nu )}
\def\LNU{L(\NU )}
\def\Lin#1{L^{-1}(#1)}
\def\L#1{L(#1)}
\def\Li#1{L_i(#1)}
\def\FNU#1{F_\NU(#1)}
\def\FinNU#1{F^{-1}_\NU(#1)}

\def\ftthetax{\tilde{f}_{\Theta|X}(\theta|x)}
\def\ER{R}

\def\Ndd#1#2#3{{\cal N}\left( #1 ;~ #2, #3 \right)}
\def\Cdd#1#2#3{{\cal C}\left( #1 ;~ #2, #3 \right)}
\def\Sdd#1#2#3#4{{\cal S}\left( #1 ;~ #2, #3, #4 \right)}
\def\Edd#1#2{{\cal E}\left( #1 ;~ #2 \right)}
\def\DEdd#1#2{{\cal DE}\left( #1 ;~ #2 \right)}
\def\Gdd#1#2#3{{\cal G}\left( #1 ;~ #2, #3 \right)}
\def\IGdd#1#2#3{{\cal IG}\left( #1 ;~ #2, #3 \right)}

\def\Xwr{X(\omega,\rb)}
\def\xwr{x(\omega,\rb)}

\def\muzw{\mu_{z}(\omega)}
\def\muzwr{\mu_{z(\rb)}(\omega)}
\def\szw{\sigma_{z}^2(\omega)}
\def\szwr{\sigma_{z(\rb)}^2(\omega)}
\def\pzw{p_{z}(\omega)}
\def\pzwr{p_{z(\rb)}(\omega)}

\def\hszw{\widehat{\sigma_{z}^2}(\omega)}
\def\hmuzw{\widehat{\mu}_{z}(\omega)}
\def\hpzw{\widehat{p}_{z}(\omega)}

\def\muzwrl{\mu_{z(\rb)}(\omega-l)}

\def\pxwrcz{p\left(\xwr ~|~ z\right)}
\def\pxwr{p\left(\xwr\right)}

\def\Xbw{\Xb(\omega)}
\def\xbw{\xb(\omega)}
\def\uXb{\underline{\Xb}}
\def\uxb{\underline{\xb}}
\def\pxbw{p(\xbw)}
\def\puxb{p(\uxb)}
\def\pzb{P(\zb)}

\def\Szw{S_z(\omega)}
\def\mubw{\mub(\omega)}

\def\Srw{S_{\rb}(\omega)}
\def\Swr{S_{\omega}(\rb)}
\def\Sbw{\Sb(\omega)}
\def\sbw{\sb(\omega)}
\def\Sbr{\Sb(\rb)}
\def\mubw{\mub(\omega)}
\def\cov#1{\mbox{cov}[#1]}

\def\pdf{\emph{pdf}}
\def\pmf{\emph{pmf}}
\def\dpdx#1#2{\frac{\partial #1}{\partial #2}}

\def\REM#1{}

\def\XXwr{X(\omega,\rb)}
\def\xxwr{x(\omega,\rb)}

\def\Xwr{X_{\omega}(\rb)}
\def\xwr{x_{\omega}(\rb)}
\def\Xrw{X_{\rb}(\omega)}
\def\xrw{x_{\rb}(\omega)}

\def\Ywr{Y_{\omega}(\rb)}
\def\ywr{y_{\omega}(\rb)}
\def\Yrw{Y_{\rb}(\omega)}
\def\yrw{y_{\rb}(\omega)}

\def\Ewr{E_{\omega}(\rb)}
\def\ewr{\epsilon_{\omega}(\rb)}
\def\Erw{E_{\rb}(\omega)}
\def\erw{\epsilon_{\rb}(\omega)}

\def\Xbr{\Xb(\rb)}
\def\xbr{\xb(\rb)}
\def\Xbw{\Xb(\omega)}
\def\xbw{\xb(\omega)}

\def\uXb{\underline{\Xb}}
\def\uxb{\underline{\xb}}
\def\uXbz{\underline{\Xb}_z}
\def\uxbz{\underline{\xb}_z}

\def\uthetab{\underline{\thetab}}
\def\uthetabz{\underline{\thetab}_z}

\def\Ywr{Y_{\omega}(\rb)}
\def\ywr{y_{\omega}(\rb)}
\def\Yrw{Y_{\rb}(\omega)}
\def\yrw{y_{\rb}(\omega)}
\def\Ybr{\Yb(\rb)}
\def\ybr{\yb(\rb)}
\def\Ybw{\Yb(\omega)}
\def\ybw{\yb(\omega)}
\def\uYb{\underline{\Yb}}
\def\uyb{\underline{\yb}}
\def\uYbz{\underline{\Yb}_z}
\def\uybz{\underline{\yb}_z}

\def\Ewr{E_{\omega}(\rb)}
\def\ewr{\epsilon_{\omega}(\rb)}
\def\Erw{E_{\rb}(\omega)}
\def\erw{\epsilon_{\rb}(\omega)}
\def\Ebr{Eb(\rb)}
\def\ebr{\epsilonb(\rb)}
\def\Ebw{Eb(\omega)}
\def\ebw{\epsilonb(\omega)}
\def\uEb{\underline{\Eb}}

\def\Erwp{E_{\rb}(\omega')}
\def\Xrwp{X_{\rb}(\omega')}

\def\xwmr{x_{\omega-1}(\rb)}
\def\muzwm{\mu_{z}(\omega-1)}

\def\rhoz{\rho_z}
\def\sez{\sigma_{\eta_z}^2}

\def\xwrp{x_{\omega}(\rb')}
\def\xrwp{x_{\rb}(\omega')}
\def\xwpr{x_{\omega'}(\rb)}
\def\xwprp{x_{\omega'}(\rb')}
\def\Xwrp{x_{\omega}(\rb')}
\def\Xwrp{X_{\omega}(\rb')}
\def\Xwpr{X_{\omega'}(\rb)}
\def\Xwprp{X_{\omega'}(\rb')}

\def\sigmaed{\sigmae^2}
\def\sigmaew{\sigmae(\omega)}
\def\sigmaedw{\sigmaed(\omega)}

\def\Sigmabe{\Sigmab_{\epsilon}}
\def\sigmabh{\widehat{\Sigmab}}
\def\xbh{\widehat{\xb}}
\def\zbh{\widehat{\zb}}
\def\qbh{\widehat{\qb}}
\def\sbh{\widehat{\sb}}

\def\argmax#1#2{\arg\max_{#1}\left\{#2\right\}}
\def\argmin#1#2{\arg\min_{#1}\left\{#2\right\}}
\def\espx#1#2{\mbox{E}_{#1}\left\{#2\right\}}
\def\esp#1{\mbox{E}\left\{#1\right\}}
\def\d#1{\mbox{~d}#1}

\def\Tr#1{\mbox{Tr}\left[#1\right]}

\def\pmatrix#1{\left[\barr{c} #1 \earr\right]}
\def\Bf{\bm{B}}
\def\gbh{\widehat{\gb}}
\def\dbh{\widehat{\db}}
\def\Tr#1{\mbox{Tr}\left[#1\right]}
\def\Cov#1{\mbox{Cov}\left[#1\right]}
\def\iid{i.i.d.~}

\def\cro#1{\left[#1\right]}
\def\acc#1{\left\{#1\right\}}
\def\aprio{\emph{a priori~}}
\def\apost{\emph{a posteriori~}}
\def\hbh{\widehat{\hb}}
\def\qh{\widehat{q}}
\def\fbt{\widetilde{\fb}}
\def\thetabt{\widetilde{\thetab}}

\def\thetae{\theta_{\epsilon}}
\def\thetaei{\theta_{\epsilon_i}}
\def\etat{\widetilde{\eta}}
\def\nubt{\widetilde{\nub}}
\def\ubt{\widetilde{\ub}}
\def\phibt{\widetilde{\phib}}
\def\lnf#1{\ln\left(#1\right)}
\def\aez{a_{\epsilon_0}}
\def\aez{b_{\epsilon_0}}
\def\Sigmah{\hat{\Sigma}}
\def\bthetae{\bar{\theta}_{\epsilon}}
\def\myexp#1{\left<#1\right>}
\def\Qfz#1{q\left(f(#1)|z(#1)\right)}
\def\Qfzb{q\left(\fb|\zb\right)}
\def\Qz#1{q\left(z(#1)\right)}
\def\Qzb{q\left(\zb\right)}
\def\Qthb{q\left(\thetab\right)}
\def\Qthe{q(\thetae)}
\def\Qmb{q\left(\mb\right)}
\def\Qvb{q\left(\vb\right)}
\renewcommand{\baselinestretch}{0.85}

\title{Joint Image Restoration and Segmentation using Gauss-Markov-Potts Prior Models and Variational Bayesian Computation: Technical Details}

\name{Hacheme~AYASSO, and Ali~MOHAMMAD-DJAFARI}

\address{Laboratoire des Signaux et Syst\`emes, \\ 
Unit\'e mixte de recherche 8506 (CNRS-SUPELEC-Univ Paris-Sud) \\  
Sup\'elec, Plateau de Moulon, 3 rue Joliot Curie, 91192 Gif-sur-Yvette, 
France.}

\begin{document}
\maketitle

\begin{abstract}
{
We propose a method to restore and to segment simultaneously images degraded by a known point spread function (PSF) and additive white noise. For this purpose, we propose a joint Bayesian estimation framework, where a family of non-homogeneous Gauss-Markov fields with Potts region labels models are chosen to serve as priors for images. Since neither the joint maximum a posteriori estimator nor posterior mean one are tractable, the joint posterior law of the image, its segmentation and all the hyper-parameters, is approximated by a separable probability laws using the Variational Bayes technique.  This yields a known probability laws  of the posterior with mutually dependent shaping parameter, which aims to enhance the convergence speed of the estimator compared to stochastic sampling based estimator. The main work is description is given in \cite{ayasso2009a}, while technical details of the variational calculations are presented in the current paper.\\
}
\end{abstract}

\begin{keywords}
 Image Restoration, Image Segmentation, Bayes procedures, Variational Bayes Approximation.
\end{keywords}

\section{\uppercase{Introduction}}
\label{sec:introduction}

A simple direct model of image restoration problem is
\beq 
\gb = \Hb \fb + \epsilonb
\label{eq01}
\eeq
where $\gb$ is the observed image, $\Hb$is a huge matrix whose elements are determined by a known point spread function, $\fb$ is the unknown image, and $\epsilonb$ is the measurement error.

In a Bayesian framework for such an inverse problem, one starts by writing the expression of the posterior law:
\beq
p(\fb|\thetab,\gb;\Mc)=
\frac{p(\gb|\fb,\thetae;\Mc) \; p(\fb|\thetab_2;\Mc)}{p(\gb|\thetab;\Mc)} 
\label{eq02}
\eeq
where $p(\gb|\fb,\thetae;\Mc)=\Nc(\Hb\fb,{\thetae}^{-1}\Ib)$, called the \emph{likelihood}, is obtained using the forward model (\ref{eq01}) and the assigned probability law $p_{\epsilon}(\epsilonb)$ of the errors, $p(\fb|\thetab_2;\Mc)$ is the assigned prior law for the unknown image $\fb$ and 
\beq
p(\gb|\thetab;\Mc)=\intg p(\gb|\fb,\thetae;\Mc) \; p(\fb|\thetab_2;\Mc) \d{\fb}. 
\label{eq03}
\eeq
is the evidence of the model $\Mc$ with hyperparameters $\thetab=(\thetae,\thetab_2)$. 

\section{Proposed prior models}
As we introduced in the previous section, the main assumption here is the piecewise homogeneity of the restored image. This model corresponds to a great number of applications where the studied image is composed of finite number of materials. We consider two prior models for the unknown image: mixture of independent Gaussians (MIG) and Mixture of Gauss-Markov (MGM).

\noindent{\bf Case 1: Mixture of Independent Gaussians (MIG):} \\ 
\begin{equation}
\left\{\barr{lcl}
p(f(\rb)|z(\rb)=k)    &=&\Nc(m_k,v_k), \quad \forall \rb\in\Rc \\ 
p(\fb|\zb,m_z,v_z)  &=&\prod_{\rb\in\Rc} \Nc(m_z(\rb),v_z(\rb))  
\earr\right.
\end{equation}
with $m_z(\rb)=m_k, \forall \rb\in\Rc_k$, $v_z(\rb)=v_k, \forall \rb\in\Rc_k$, and  

\noindent{\bf Case 2: Mixture of Gauss-Markovs(MGM):}\\ 
\begin{footnotesize}
\begin{equation}
p(f(\rb)|z(\rb)=k,f(\rb'),z(\rb'),\rb'\in\Vc(\rb))=\Nc(\mu_k(\rb),v_k(\rb))
\end{equation}
\end{footnotesize}
with

\begin{footnotesize}\beq 
\left\{\barr{lcl}
\Cc(\rb) &=& 1-\prod_{\rb'\in\Vc(\rb)} \delta(z(\rb')-z(\rb))\\
\mu_k(\rb)&=&\left\{\barr{lll}
m_k & \mbox{if} &  \Cc(\rb)=1 \\ 
\frac{1}{|\Vc(\rb)|}\sum_{\rb'\in\Vc(\rb)} f(\rb')   & \mbox{if} & \Cc(\rb)=0 
\earr\right.
\\
v_k(\rb)&=&v_k\qquad \forall \rb\in\Rc_k
\earr\right.
\label{eq14}
\eeq
\end{footnotesize}

For the hidden field $\zb$, a Potts Markov model will be used to describe the hidden field prior law for both image models:
\begin{small}\beq 
\barr{l@{}l@{}l}
p(\zb|\gamma)  &\propto&
 \exp  \left[  \sum_{\rb \in \Rc}  \Phi(z(\rb))+\frac{1}{2}\gamma  \sum_{\rb\in\Rc}\sum_{,\rb'\in\Vc(\rb)} \delta(z(\rb)-z(\rb')) \right]
\earr
\label{eq15}
\eeq\end{small}
where $\Phi(z(\rb))$ is the energy of singleton cliques, and $\gamma$ is Potts constant. The hyperparameters of the model are class means $m_k$, variances $v_k$, and finally singleton clique energy $\kappa_{k} = \Phi(k)$.

For the hyperparameters $\thetab=\{\thetae,\mb,\vb,\kappab \}$, we choose the following prior laws: inverse Gamma for the model of the error variance $\thetae$, Gaussian for the means  $m_k$, Inverse Gamma for variances $v_k$, and finally a Dirichlet for $\kappa_k$.

\begin{eqnarray}
p(\thetae|\alpha_{0},\beta_{0}) &=&\Gc(\alpha_{0},\beta_{0}), \; \forall k \\ 
p(m_k|m_0,\sigma_0)         &=&\Nc(m_0,\sigma_0), \; \forall k \\ 
p(v^{-1}_k|a_0,b_0)&=&\Gc(a_0,b_0), \; \forall k \\ 
p(\kappab|\kappa_0)    &=&\Dc(\kappa_0,\cdots,\kappa_0) 
\end{eqnarray}
where $\alpha_{0},\beta_{0}$, $m_0,\sigma_0$, $a_0,b_0$ and $\kappa_0$ are fixed for a given problem.

\section{Variational Bayes Approximation}
In Bayesian framework, the joint estimation of the image $\fb$, the segmentation $\zb$, and the hyperparameter $\thetab$ can be done from the joint posterior, 
\beq
\barr{lll}
p(\fb,\zb,\thetab|\gb;\Mc)&\propto &p(\gb|\fb,\thetae;\Mc) \; p(\fb|\zb,\mb,\vb;\Mc)\;p(\zb|\kappab,\gamma;\Mc)\; p(\thetab|\Mc)
\earr
\eeq
One of the main difficulties to obtain an analytical estimator is the posterior dependence between the searched parameters. For this reason, we propose, in this kind of methods, a separable form of the joint posterior law $q$, and then we try to find the closest posterior to the original posterior under this constraint in terms of the Kullback-Leibler divergence:  
\beq
\barr{lll}
\mbox{KL}(q:p)&=& \displaystyle{\int} q(\xb) \ln \frac{q(\xb)}{p(\xb|\Mc)} \d{\xb} \\ 
              &=& -\sum_j  H(q_j) - \left< \ln p(\xb|\Mc)\right>_{q(\xb)}
\earr 
\eeq
So, the main mathematical problem to study is finding $\qh(\xb)$ which minimizes $\mbox{KL}(q:p)$. 
Using property of the exponential family, this functional optimization problem can be solved and we obtain: \\
\beq
q_j(x_j) = \frac{1}{C_j} \expf{-\left< \ln p(\xb|\Mc)\right>_{q_{-j}}}
\eeq 
where $q_{-j}=\prod_{i\not=j} q_i(x_i)$ and $C_j$ are the normalizing factors. 
Choosing a strongly separated posterior, where only dependence between image pixels and hidden fields is conserved:
\begin{small}\beq
q(\fb,\zb,\thetab)=\prod_{\rb} \left[q(f(\rb)|z(\rb))\right] \prod_{\rb}\left[q(z(\rb))\right] \; \prod_{l}q(\thetab_l)
\eeq\end{small}
and applying this approximated expression for $p(\fb,\zb,\thetab|\gb;\Mc)$, we obtain:
\begin{eqnarray}
q(\fb|\zb)& = &\prod_{k}\prod_{\rb\in\Rc_k} \Nc(\tilde{\mu}_k(\rb),\tilde{v}_k(\rb))\\
q(\zb) &=& \prod_{\rb} p(z(\rb)|\tilde{z}(\rb'), \rb'\in\Vc(\rb)) \\
q\scriptstyle{({z(\rb)=k}|\tilde{z}(\rb'))} &=&\tilde{\zeta}_{k}(\rb) \propto \tilde{c}_k \; \tilde{d}_k(\rb) \; \tilde{e}_k(\rb) \\
q(\thetae|\tilde{\alpha},\tilde{\beta})  &=&\Gc(\tilde{\alpha},\tilde{\beta}) \\ 
q(m_k|\tilde{m}_k,\tilde{\sigma}_k)       
&=&\Nc(\tilde{m}_k,\tilde{\sigma}_k), \; \forall k \\ 
q(v^{-1}_k|\tilde{a}_k,\tilde{b}_k)  
&=&\Gc(\tilde{a}_k,\tilde{b}_k), \; \forall k \\ 
q(\kappab)&=&\Dc(\tilde{\kappa}_1,\cdots,\tilde{\kappa}_K)
\label{eq23}
\end{eqnarray}
where tilded values need to be estimated in an iterative way since they are mutually dependent. We give in the following the expression of these values in the iteration $t$ as function other values in the iteration $t-1$. 

\noindent We start by $\tilde{\mu}_k^t$ and $\tilde{v}_k^t$
\begin{eqnarray}
\tilde{\mu}_k^{t}(\rb) &{}={}& \tilde{f}^{t-1}(\rb) + \tilde{v}_k^{t}(\rb) \Biggl[  \frac{\left(\tilde{\mu}_{k}^{*t-1}(\rb)-\tilde{f}^{t-1}(\rb)\right)}{\bar{v}_{k}^{t-1}} \Biggr.
\Biggl. {\bthetae}^{t-1} \sum_{\sb} H(\sb,\rb) \left( g(\sb) - \tilde{g}^{t-1}(\sb) \right) \Biggr],\label{eq:sys4_eq_1}\\
\tilde{\mu}^{*t}_{k}(\rb) &{}={}& \left\{ 
\begin{array}{ll}
\tilde{m}^{t}_k &\mbox{MIG case} \\ 
 \frac{1-\tilde{\Cc}_k^{t}(\rb)}{|\Vc(\rb)|}\sum_{\rb'\in\Vc(\rb)} \tilde{\mu}_{k}^{t}(\rb')
+ \tilde{\Cc}_k^{t}(\rb) \tilde{m}_k^{t}, &\mbox{MGM case} \label{eq:sys4_eq_2}
\end{array}
\right.\\
\tilde{\Cc}_k^{t}(\rb) &{}={}& \left\{ 
\begin{array}{ll}
  1 & \mbox{MIG case} \\
  1- \prod_{\rb'\in\Vc(\rb)} \tilde{\zeta}_k(\rb'), &\mbox{MGM case} \label{eq:sys4_eq_3}
\end{array}  
\label{eq25}
\right.\\
\tilde{v}_k^{t}(\rb) &{}={}& \frac{\bar{v}_{k}^{t-1}}{\left( 1 + \bar{v}_{k}^{t-1} \bthetae^{t-1} \sum_{\sb} H^{2}(\sb,\rb) \right)},\label{eq:sys5_eq_1}\\
\bar{v}_{k}^{t-1} &{}={}& \left< v_{k} \right>_{q^{t-1}}  = (\tilde{a}^{t-1}_k \tilde{b}_k^{t-1})^{-1},\label{eq:sys5_eq_2}\\
{\bthetae}^{t-1} &{}={}& \left< \thetae \right>_{q^{t-1}}  = (\tilde{\alpha}^{t-1} \tilde{\beta}^{t-1})^{-1},\label{eq:sys5_eq_5}\\
\tilde{f}^{t}(\rb) &{}={}& \sum_k \tilde{\zeta}_k^{t}(\rb) \tilde{\mu}_k^{t}(\rb),\label{eq:sys5_eq_3}\\
\tilde{g}^{t}(\sb)&{}={}&\left[\Hc\fb\right](\sb)=\sum_{\rb}H(\sb,\rb)\;\tilde{f}^{t}(\rb).\label{eq:sys5_eq_4}
\end{eqnarray}
While $\tilde{c}_k^t,\, \tilde{d}_k^t$ and $\tilde{e}_k^t$ are given as 
\begin{eqnarray}
\tilde{c}_k^{t} &{}={}& \exp\biggl[\Psi(\tilde{\kappa}_{k}^{t-1})-\Psi(\sum_{l}\tilde{\kappa}_{l}^{t-1}) 
{+}\: \frac{1}{2}\left(\Psi(\tilde{b}_k^{t-1})+\lnf{\tilde{a}_{k}^{t-1}}\right)\biggr],\label{eq:sys6_eq_1}\\
\tilde{d}_k^{t}(\rb) &{}={}& \sqrt{(\tilde{v}^{t}_k(\rb))^{-1}}\:\mbox{exp}\left[-\frac{1}{2} \left(\frac{\left[\tilde{\mu}_k^{t}(\rb)-\tilde{\mu}_k^{t-1}(\rb)\right]^2}{\tilde{v}_k^{t}(\rb)} 
{-}\:
\frac{\left(\tilde{\mu}^{t}_k(\rb)\right)^2}{\tilde{v}^{t}_k(\rb)}+\frac{\myexp{\left(\mu_k(\rb)\right)^2}_{q^{t-1}}}{\bar{v}^{t-1}_k} \right)  \right],\label{eq:sys6_eq_2}\\ 
\tilde{e}_k^{t}(\rb)&{}={}& \expf{+\frac{1}{2}\gamma \sum_{\rb'}\tilde{\zeta}_k^{t-1}(\rb')}. \label{eq:sys6_eq_3}
\end{eqnarray}
Finally, the rest of the shaping parameters are 
\begin{eqnarray}
\tilde{\alpha}^{t} &{}={}& \left[ \alpha^{-1}_{0} + \frac{1}{2} \sum_{\rb}\left<(g(\rb)-\left[\Hc\fb\right](\rb))^2\right>_{q^{t-1}}\right]^{-1},\label{eq:sys7_eq_1}\\
\tilde{\beta}^{t}  &{}={}& b_{0}+  \frac{\left|\Rc\right|}{2},\label{eq:sys7_eq_2}\\
\tilde{m}_k^{t} &{}={}& \tilde{\sigma}_{k}^{t} \left( \frac{m_0}{\sigma_0}+\frac{1}{\bar{v}_{k}^{t-1}} \sum_{\rb} \tilde{\Cc}_k^{t-1}(\rb)\tilde{\zeta}_{k}^{t-1}(\rb) \tilde{\mu}_{k}^{t-1}(\rb) \right),\label{eq:sys7_eq_3}\\
\tilde{\sigma}_k^{t} &{}={}& \left( \sigma_0^{-1} +\frac{1}{\bar{v}_{k}^{t-1}} \sum_{\rb} \tilde{\Cc}_k^{t-1}(\rb)\tilde{\zeta}_{k}^{t-1}(\rb)\right)^{-1}, \label{eq:sys7_eq_4}\\
\tilde{a}_k^{t} &{}={}&  \left[ a^{-1}_0 + \frac{1}{2} \sum_{\rb} \left< \left(f(\rb)-\mu_k(\rb)\right)^2\right>_{q^{t-1}} \right]^{-1}, \label{eq:sys7_eq_5}\\
\tilde{b}_k^{t} &{}={}& b_0 + \frac{1}{2} \sum_{\rb} \tilde{\zeta}_k^{t-1}(\rb),\label{eq:sys7_eq_6}\\
\tilde{\kappa}_k^{t}&{}={}& \kappa_{0} + \sum_{\rb} \tilde{\zeta}^{t-1}_{k}(\rb).\label{eq:sys7_eq_7}
\end{eqnarray}
with
\begin{eqnarray}
\myexp{\left(g(\sb)-\sum_{\rb}H(\sb,\rb)f(\rb)\right)^2}_{q_{-\thetae}} &{}={}& g^{2}(\sb) -2g(\sb)\tilde{g}(\sb)+(\tilde{g}(\sb))^2\nonumber\\
&&{+}\:\sum_{\rb}H^2(\sb,\rb)\left[\sum_k\tilde{\zeta}_{k}(\rb)\left((\tilde{\mu}_k)^2(\rb)+\tilde{v}_k(\rb)\right)-\tilde{f}^2(\rb)\right]
\end{eqnarray}
and
\begin{eqnarray}
\myexp{\left(f(\rb)-\mu_k(\rb)\right)^2}_{q_{-\vb}} &{}={}& \tilde{\zeta}_{k}(\rb)\biggl[\left(\tilde{\mu}_k(\rb)\right)^2+\tilde{v}_k(\rb)-2\tilde{\mu}_k(\rb)\tilde{\mu}_k^{*}(\rb)\biggr.\nonumber\\
&&{+}\:\biggl.\frac{1-\tilde{\Cc}_k(\rb)}{|\Vc(\rb)|}\sum_{\rb'\in\Vc(\rb)}\left[ \left(\tilde{\mu}_k(\rb')\right)^{2}+\tilde{v}_k(\rb)\right]+\tilde{\Cc}_k(\rb) \left[ \left(\tilde{m}_k\right)^2+\tilde{\sigma}_k\right]\biggr]
\end{eqnarray}

\section{Conclusion}
A variational Bayes approximation is proposed in this paper for image restoration. We have introduced a hidden variable to give a more accurate prior model of the unknown image. Two priors, independent Gaussian and Gauss-Markov models were studied with Potts prior on the hidden field. Technical Details of this approximation were presented .

\bibliographystyle{ieeetr}
\def\UP#1{\uppercase{#1}}

\end{document}